\documentclass[10pt, conference]{IEEEtran}
\IEEEoverridecommandlockouts

\usepackage{cite}
\usepackage{amsmath,amssymb,amsfonts}
\usepackage{algorithmic}
\usepackage{graphicx}
\usepackage{textcomp}
\usepackage{xcolor}
\def\BibTeX{{\rm B\kern-.05em{\sc i\kern-.025em b}\kern-.08em
    T\kern-.1667em\lower.7ex\hbox{E}\kern-.125emX}}
\begin{document}

\title{
Managing Cold-start in The Serverless Cloud with Temporal Convolutional Networks
}

\author{\IEEEauthorblockN{Tam N. Nguyen}
\IEEEauthorblockA{
tom.nguyen@ieee.org\\
linkedin.com/in/tamcs}
}

\maketitle

\begin{abstract}
Serverless cloud is an innovative cloud service model that frees customers from most cloud management duties. It also offers the same advantages as other cloud models but at much lower costs. As a result, the serverless cloud has been increasingly employed in high-impact areas such as system security, banking, and health care. A big threat to the serverless cloud's performance is cold-start, which is when the time of provisioning the needed cloud resource to serve customers' requests incurs unacceptable costs to the service providers and/or the customers. This paper proposes a novel low-coupling, high-cohesion ensemble policy that addresses the cold-start problem at infrastructure- and function-levels of the serverless cloud stack, while the state of the art policies have a more narrowed focus. This ensemble policy anchors on the prediction of function instance arrivals, 10 to 15 minutes into the future. It is achievable by using the temporal convolutional network (TCN) deep-learning method. Bench-marking results on a real-world dataset from a large-scale serverless cloud provider show that TCN out-performs other popular machine learning algorithms for time series. Going beyond cold-start management, the proposed policy and publicly available codes can be adopted in solving other cloud problems such as optimizing the provisioning of virtual software-defined network assets.
\end{abstract}


\section{Introduction}
Cloud computing is on the rise as it improves three strategic business objectives of flexibility, efficiency, and innovation. Businesses can scale their cloud infrastructure on demand, enjoy secure pre-built tools, benefit from highly reliable storage options, while being able to off-load certain responsibilities to the cloud service providers (CSPs). Cloud computing is also efficient as it offers a wider range of accessibility, a faster speed to market, more consistent data security, more savings on equipment, and most importantly, more flexible pay structures. Therefore, businesses can innovate and monetize their innovations faster.

Managing cloud systems involves challenges with load balancing, auto-scaling, cyber security, monitoring, and so on \cite{Shahrad2020ServerlessProvider}. To relieve customers from cloud management duties, Amazon first introduced Amazon Lambda in 2014 after which big CSPs like Google and Microsoft adopted in 2016. Amazon Lambda is an example of "serverless cloud service" or "Function as a Service" (FaaS) which offers a pay-as-you-go model, and service auto-scale feature \cite{Shafiei2022ServerlessApplications}. Customers only need to register their functions with the CSPs. Upon invocations, cloud service resources will be assigned to successfully execute the functions. Customers will receive the results without worrying about behind-the-scene cloud logistics.

Compared to other cloud service models, the serverless model is more affordable due to resource multiplexing and infrastructure heterogeneity \cite{Shafiei2022ServerlessApplications}. For example, Deloitte \cite{Tayal2019DeterminingConsulting} reported that Amazon serverless offers the same capabilities as Amazon EC2 but only with 65 percent of the cost. With such wonderful features, the serverless clouds have been increasingly employed in high-impact areas such as Real-time Collaboration and Analytic, Urban and Industrial Management, System and Software Security, E-commerce, Health Care, Banking, and Blockchains. Therefore, maintaining a high performance level of the serverless clouds has became more and more critical to both CSPs and their customers.

A big threat to serverless cloud's performance is cold-start, which is when the time of provisioning the needed cloud resource to serve customers' requests incurs unacceptable costs to the service providers and/or the customers. Solving this problem is not trivial as making the resources available at all times. For the main reason, idling resources must be un-provisioned to keep the costs low.

From reviewing literature, the paper first identifies five real-life key challenges in managing cold-start and current gaps in addressing those challenges (section \ref{sec_background}). To assist current and future cold-start management efforts with machine learning, the five challenges were operationalized into two specific sets of data requirements (DRs) and prediction model requirements (MRs). From the Microsoft (MS) Azure Functions Trace 2019 dataset \cite{Shahrad2020ServerlessProvider}, the paper produces a training data set that satisfies all DRs (section \ref{sec_dataset}). In section \ref{sec_design}, a brief discussion on mainstream algorithms for time series forecasting sets the stage for Temporal Convolutional Networks as the algorithm of choice in satisfying all MRs. The deep learning pipeline has two chained modules to predict future instances' function names (module A) and predict those instances' arrival time (module B).

Anchors on the capability of predicting function instance arrivals 10 to 15 minutes into the future (section \ref{sec_results}), this paper proposes a novel low-coupling, high-cohesion ensemble policy that addresses the cold-start problem at both infrastructure- and function-levels of the serverless cloud stack (section \ref{sec_policy}). Going beyond cold-start management, the proposed policy together with the paper's publicly shared source codes can be adopted in solving other cloud problems such as optimizing the provisioning of virtual software-defined network assets.

\begin{table*}[htbp]
\caption{Recent Works on Temporal Predictions of FaaS Workloads}
\begin{center}
\begin{tabular}{|l|l|l|l|c|c|c|c|c|}
\hline
\multicolumn{1}{|c|}{\textbf{Works}} & \multicolumn{1}{c|}{\textbf{Year}} & \multicolumn{1}{c|}{\textbf{Prediction   Target}} & \multicolumn{1}{c|}{\textbf{Method}} & \multicolumn{1}{c|}{\textbf{C1}} & \multicolumn{1}{c|}{\textbf{C2}} & \multicolumn{1}{c|}{\textbf{C3}} & \multicolumn{1}{c|}{\textbf{C4}} & \multicolumn{1}{c|}{\textbf{C5}} \\ 
\hline
Mahmoudi et. al. \cite{Mahmoudi2020PerformancePlatforms} & 2022 & Cold-start probability & Semi-Markov process &  &  & $\bullet$ & $\bullet$ & $\bullet$ \\ 
\hline
Ebrahimpour et. al. \cite{Ebrahimpour2022AEnvironments} & 2022 & Function merging time & Heuristic method on dependency graph &  & $\bullet$ & $\bullet$ &  & $\bullet$ \\ 
\hline
Jegannathan et. al. \cite{Jegannathan2022APlatform} & 2022 & Workload arrival time & SARIMA &  & $\bullet$ &  & $\bullet$ & $\bullet$ \\ 
\hline
Steinbach et. al. \cite{Steinbach2022TppFaaS:Processes} & 2022 & Function invocations & Neural temporal point processes &  & $\bullet$ & $\bullet$ & $\bullet$ & $\bullet$ \\ 
\hline
Singhvi et. al. \cite{Singhvi2021Atoll:Platform} & 2021 & Workload arrival rate & Exponentially weighted moving average &  &  & $\bullet$ & $\bullet$ & $\bullet$ \\ 
\hline
Lin et al. \cite{Lin2021ModelingApplications} & 2021 & Workflow response time & Critical path greedy   algorithms &  & $\bullet$ & $\bullet$ &  & $\bullet$ \\ 
\hline
Shahrad et. al. \cite{Shahrad2020ServerlessProvider} & 2020 & Function workload & Histogram analysis and Arima &  & $\bullet$ & $\bullet$ &  & $\bullet$ \\ 
\hline
Anirban Das et. al. \cite{Das2020PerformancePlacement} & 2020 & End-to-end latencies & Gradient boosting regressor &  &  &  & $\bullet$ & $\bullet$ \\
\hline
\multicolumn{9}{l}{$^{\mathrm{a}}$C1 to C5 are the previously stated real-world challenges of solving cold-start.}
\end{tabular}
\label{tab:my-table}
\end{center}
\end{table*}

\section{Background}\label{sec_background}
\begin{figure}[htbp]
\centerline{\includegraphics[width=2.3in]{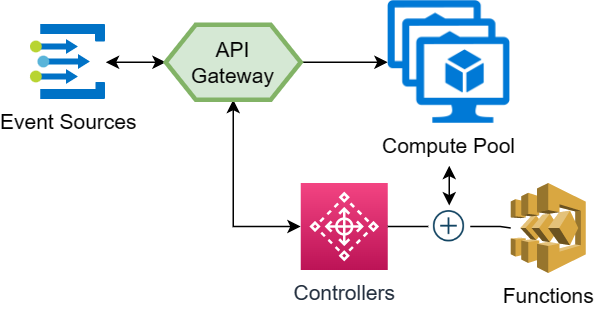}} 
\caption{FaaS Workflow}
\label{fig_faas_workflow}
\end{figure}

A serverless service can be either FaaS or Backend as a Service (BaaS) \cite{Shafiei2022ServerlessApplications}. This paper focuses on FaaS which consists of a client-maintained application and its registered functions with a CSP. The functions can be triggered via HTTP requests, timer events, command line interfaces, pushed updates to a message queue, or other mechanisms. Upon trigger receipts, an API gateway performs necessary checks and maps the triggers with correct functions, per clients' predefined rules. The controller cooperates with the API gateway and determines which worker node in the compute pool will receive which function for execution. The controller usually has three main components of a resource scheduler, a resource scaler, and a resource monitor \cite{Mampage2022ADirections}. A worker node is a lean, isolated virtual environment with only needed resources and configurations for successful function execution. Once the function was successfully executed, the worker will send the results to the controller. The controller then forward the results to the API gateway to be sent back to the customer. The controller may also shutdown the worker node(s) to conserve resources. Figure \ref{fig_faas_workflow} summarizes the FaaS workflow.

Notably, the creation of a worker node involves initializing a new container, setting up code dependencies, configuring the right run-time environment, and other requirements \cite{Agarwal2021AFrequency}. Node provisioning is dependent on autoscaling strategies such as scale-per-request, concurrency value scaling, or metric-based scaling \cite{Mahmoudi2020PerformancePlatforms}. Most CSPs follow the scale-per-request strategy, which involves provisioning a new worker node when there is no available provisioned worker node (a.k.a "warm node") to handle an arrived function request \cite{Mahmoudi2020PerformancePlatforms}. When the total execution time of a customer's serverless function request is unacceptably long due to a new node's long provision time, we have a cold-start problem.

The cold-start problem can be solved by making worker nodes readily available. However, doing so is in conflict with the resource conservation policy that shuts down nodes. Most CSPs use a fixed keep-alive policy to sustain worker nodes for fixed amounts of time \cite{Shahrad2020ServerlessProvider}. This policy, late-binding policy, and random load-balancing policy were found not to be efficient because they fail to fully address the following real-life challenges \cite{Shahrad2020ServerlessProvider, Kaffes2022Hermod:Functions}.

\begin{itemize}
    \item The customers' needs for running FaaS applications are hard to predict. (\textbf{C1})
    \item FaaS applications are heterogeneous with widely varied function invocation patterns. (\textbf{C2})
    \item Some FaaS applications run very infrequently. (\textbf{C3})
    \item The cost of FaaS function invocation tracking should be reasonable. (\textbf{C4})
    \item The execution of FaaS function optimization policy should be fast enough. (\textbf{C5})
\end{itemize}

Effective cold-start management policies often result from well-execution of workload characterization, performance prediction, resource scheduling, and resource scaling \cite{Mampage2022ADirections}. This paper focuses on the task of predicting FaaS workloads as the anchor for the resource scheduling and the resource scaling. Recent works on predicting FaaS workloads are listed in Table \ref{tab:my-table}. For a major distinction, this paper focuses on solving the task of predicting future FaaS function instances' arrival time well by addressing all of the stated five challenges, which were operationalized as the following dataset and model requirements. 

Dataset(s) used for prediction model training and validating must:
\begin{itemize}
    \item Be derived from a real-world dataset from a major FaaS CSP. (\textbf{DR1})
    \item Involves a diverse pool of users and applications. (\textbf{DR2})
    \item Involves applications that are heterogeneous with hard-to-predict invocations. (\textbf{DR3})
    \item Involves some functions that are executed very infrequently. (\textbf{DR4})
    \item Not have obvious secular trend nor seasonal variations. (\textbf{DR5})
    \item Have complex temporal dependencies. (\textbf{DR6})
\end{itemize}

Forecasting models must:
\begin{itemize}
    \item Be competitively accurate in predicting future FaaS function arrival time. (\textbf{MR1})
    \item Support incremental learning. (\textbf{MR2})
    \item Support a practical number of steps into the future based on generalized knowledge of past steps. (\textbf{MR3})
    \item Be fast enough. (\textbf{MR4})
\end{itemize}

The next sections provide further details on dataset construction, and model design.

\section{The Dataset}\label{sec_dataset}
The paper utilizes the Azure Functions Trace 2019 dataset that was originally published by Sharad et. al. \cite{Shahrad2020ServerlessProvider}. Data was collected from a real-world Microsoft Azure Function FaaS environment (satisfying DR1 and DR2). The dataset and its full documentation are available at \textit{github.com/Azure/AzurePublicDataset}. The most relevant data fields are as follows. HashOwner contains unique IDs of the application owners. HashApp contains unique IDs for application names. HashFunction contains unique IDs for the function names within the apps. Average represents the average execution time of a function instance.

\begin{figure}[htbp]
\centerline{\includegraphics[width=3.3in]{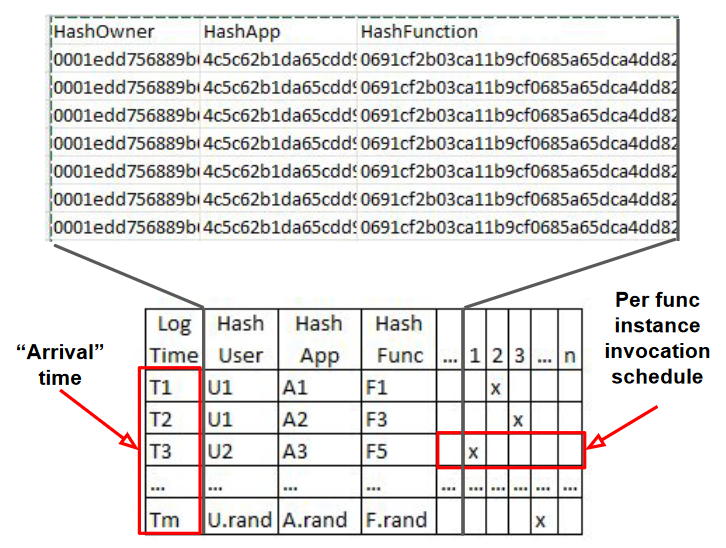}} 
\caption{Function Instance Arrival Time v.s. Function Instance Invocation Schedule}
\label{fig_arrival_vs_schedule}
\end{figure}

Function invocations of each function instance were recorded for a total duration of 14 days. Function triggers were grouped into 7 classes of HTTP, Event, Queue, Timer, Orchestration, Storage, and others. An application may encompass multiple functions. A function can be loaded multiple times into different worker nodes. Each of this loaded instance has its own invocation schedule. There are many reasons for multiple instances of the same function such as multiple function calls within or across applications written by the same owner. Figure \ref{fig_arrival_vs_schedule} shows an example of function instances' arrivals and their invocation schedules.

In this example, there are 7 instances of the same function of the same application and same user. Further inspection of the dataset shows the following invocation schedules: Instance 1 [578], Instance 2 [639, 642, 1288, 1292], Instance 3 [622, 627], Instance 4 [772, 773,776], Instance 5 [559, 566], Instance 6 [714, 723, 1100, 1102,1109], and Instance 7 [677, 679, 690,704,1098]. The numbers indicate the minute positions in the global timeline of 14 days. It is important to note that these per-instance invocation schedule is different from the function instance arrival time. While Sharad et. al. \cite{Shahrad2020ServerlessProvider} focused on solving the cold-start problem by optimizing the function instance invocation schedules, this paper focuses on optimizing the function instance processes by predicting function instance arrival time instead.  

\subsection{Data Exploration Results}
Data set exploration results including detailed charts can be found in the original paper by Shahrad et. al. \cite{Shahrad2020ServerlessProvider}. The key results can be summarized as follows.

The majority of function execution time is equal to or less than the time it takes to initialize the needed resources for function execution. Hence, cold-start is a real problem. The vast majority of applications were invoked infrequently which satisfies DR4. Less than 20\% of the applications were responsible for 99.6\% of all invocations which means there is a huge variations in function invocations among applications. Additionally, the average inter-arrival time of functions among applications varies greatly. These two findings satisfy DR3 and DR5. The least infrequent function invocations were executed 8 times less frequent than the most popular ones. This finding supports DR4. HTTP triggers play an influential role as 64\% of the applications have at least one HTTP triggered function request, and 43\% of the applications have only HTTP triggered functions \cite{Shahrad2020ServerlessProvider}. The average function execution time of different triggers also varies greatly. Memory was found to be the important factor affecting functions' warm-up, allocation, and keep-alive decisions. These two findings support DR6.

Based on the data exploration results, this paper focuses on the following objectives.
\begin{enumerate}
    \item \textbf{Predict the name of the future HTTP-triggered function instances.}
    \item \textbf{Predict the arrival time for each of those function instances.}
\end{enumerate}

\subsection{Data Preparation}
Training dataset was prepared using Azure Synapse Data Lake-House and PySpark. The original dataset has 662927 HTTP-triggered function instances. Considering the global timelne of 14 days, there are around 33 provisioned HTTP-triggered function instances per minute. The filtered and exported dataset has 136497 rows. The main data fields are Func\_index (function instance index), HashOwner, HashApp, Func\_ID (the numerical mappings of function instance names), and Average (the average execution time of each function instance). Data preparation Jupyter Notebooks with descriptions for each data manipulation step and exported datasets can be found at \textit{https://github.com/Cybonto/TCN4CLOUD}. The key data manipulation decisions can be briefly summarized as follows.

Only HTTP-triggered function instances were selected because HTTP-related invocations are the most unpredictable and most popular type of invocations within the original Azure FaaS dataset. Then, only functions that have more than 9 provisioned instances were selected. Each function instance has its own function invocation schedule. Selecting this filter does not sacrifice the complexities of having functions with less than 9 and more than 9 provisioned instances within the same training dataset. For example, along the time line,  we still have the early instances (the 1st to 5th) of certain functions mixed with the later instances (the 9th to the 15th) of other functions. The benefits of having this filter goes beyond having efficient training data. By selecting only functions that have more than 9 instances, the time gaps and their variances among function instance arrivals increased, better satisfying DR3, DR5, and DR6. In fact, the resulting average variance of the time gaps is 3.6 time-ordered rows.

\section{Model Design}\label{sec_design}
\begin{figure}[htbp]
\centerline{\includegraphics[width=2.4in]{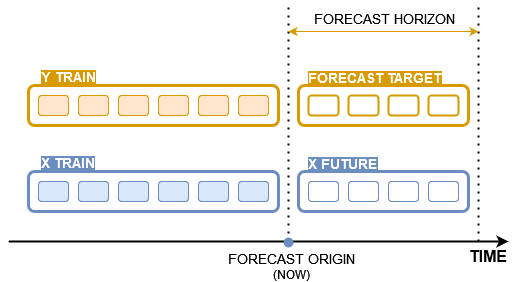}} 
\caption{Time Series Forecasting}
\label{fig_timeseries_forecasting}
\end{figure}

The time series $T$ with length $n \in \mathbb{N}$ can be defined as the set of pairs $T = \{(t_i , p_i)|t_i \leq t_{i+1} , 0 \leq i \leq n\}$ with data points $p_i \in R_d$ that have $d$ behavioural attributes and  timestamps $t_i \in \mathbb{N}, i \leq n$. When $d = 1, T$ is \textit{univariate}, and when $d > 1, T$ is multivariate. Figure \ref{fig_timeseries_forecasting} demonstrates the key concepts of time series forecasting. A time series may have characteristics of stationarity (a constant mean, variance, and auto-correlation structure), seasonality (periodically reoccurring events), trends, pulses, and so on. Based on Azure ML analysis, there is no seasonality nor specific trends in the training dataset. This fact renders traditional machine learning algorithms for time series not ideal for achieving the objectives. Instead, the paper targets deep learning algorithms which were known to have better performance \cite{Benidis2022DeepSurvey}.

The training dataset can be considered as a collection of many time series with each one corresponds to a particular function's instance arrival schedule. Training one model for each of these series will violate MR4 and may be too expensive to satisfy MR3. Hence, the deep learning model must be a global model that collectively learns from all the time series. MR3 and MR4 are most efficiently satisfied when this global model can accurately make hundreds of prediction steps into the future in one go. This number of predictions is referred to as the forecast horizon. Specifically, the model's deep learning algorithm must support direct forecasting instead of step-by-step predictions or recursive predictions. For example, the prediction of future step 2 does not require the validation of future step 1. Such model can be formalized as:
\newline
\newline
$P(z_{i,t+1:t+h} |Z_{1:t}, X_{1:t+h}; \theta_i ), \theta_i = \Psi(z_{i,1:t}, X_{i,1:t+h}, \Phi), $
\newline
\newline
We have $Z=\{z_{i,1:T_i}\}^N_{i=1}$ as a set of $N$ univariate time series with $z_{i,1:T_i}$ as the values. $\{X_{i,1:T_i}\}^N_{i=1}$ are the covariate vectors that contain both temporal and static features. $\theta_i$ is the $i^{th}$ set of model parameters. $\Phi$ are shared parameters among all N time series. $\Psi$ is the selected deep learning algorithm. This model can be implemented using deep learning architecture.

In a 2022 survey, Benidis et. al. \cite{Benidis2022DeepSurvey} listed the following deep learning algorithm families for temporal predictions together with their limitations.
\textit{Feedforward Neural Networks} (FNNs) are stacked layers of fully connected neurons. They are not the best option due to problems with fixed input/output size, with vanishing gradients, and may not fully exploit temporal features.
\textit{Convolutional Neural Networks} (CNNs) are special neural networks (NNs) with convolutional layers to exploit dimensional features commonly found in images and time series. CNNs is able to learn the causal convolutions (a.k.a the receptive field) which is essential for accurate forecasting. TCNs are the CNNs that were designed for temporal forecasting.
\textit{Recurrent Neural Networks} (RNNs) are special NNs that were designed to handle sequential data. Compared to CNNs and TCNs, RNNs are more susceptible to vanishing or exploding gradients.

\textit{Transformer} is a modern architecture that is based on the attention mechanism which can be efficient in performing long sequence prediction task. Compared with TCNs, transformers is only more efficient in training for longer prediction sequence length which is not needed for the identified objectives. 
\textit{Graph Neural Networks} (GNNs) can predict future values by learning the temporal embedding together with other embeddings. However, it is only better than TCN if there is a strategy for utilizing its graph information which is not needed for the identified objectives.
\textit{Generalized Adversarial Networks} (GANs) are being used in forecasting such as synthesizing temporal data, and measuring adversarial temporal loss. GANs are usually used with other members of different families and may over complicate the paper's model design and implementation.  

Considering the above-mentioned Benidis et. al. \cite{Benidis2022DeepSurvey} analysis of major deep learning families for temporal predictions, implementation efficiency, and the algorithms' strengths in addressing the stated model requirements (MRs), Temporal Convolutional Networks (TCNs) was selected for implementation.

\subsection{TCN Implementation}
TCNs uses historical data of function owners' HTTP-triggered function instance arrivals, the average execution time for each function instance, and the number of provisioned instances of the same function to predict what function instance will be provisioned, and their arrival time. The forecast horizon was set to 500 arrivals which equals 10 to 15 minutes into the future of the Azure FaaS operations. This amount of time should be plenty for any downstream cold-start management tasks.

In the first cross validation step (step 1), a candidate model will just observe and learn from the first 500 arrivals. In the second validation step, the window of 500 arrivals slides to the end of the first step. The model will learn from past knowledge ( 500 pairs of X and Y values in step 1) and current step's 500 X values to predict the current step's 500 Y values (the instance arrivals). Before moving to step 3, it will validate its predicted Y values with the ground-truth Y values in the train dataset for that step window. The process repeats until the process reaches the end of the train data set or when the model's performance cannot be improved.

\begin{figure}[htbp]
\centerline{\includegraphics[width=2.8in]{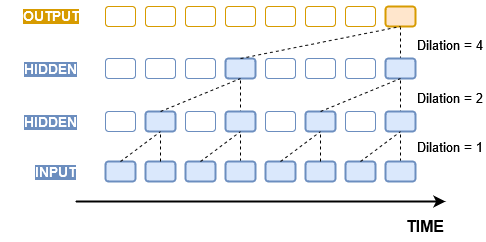}} 
\caption{The Dilation Mechanism}
\label{fig_dilation}
\end{figure}

This paper's TCN implementation has three main components. A pre-mix layer mixes the input data into an array of signals. A stack of dilated convolution layers will process the mixed input signals of which mechanism was illustrated in Figure \ref{fig_dilation}. The outputs from each layer produce a new channel array containing a mixture of dilated convolution-filtered signals. These outputs will be feed as inputs to the upper stack. Finally, the forecast layer coalesces the output signals and generates the forecasts. 

These dilated, causal convolution layers are essential to the TCNs' prediction strength. The convolution mixes and combine signals around a certain time point while the dilation decides the neighborhood size. When we stack the layers higher, one time point in the top layer will have a wider convolutionized base that covers a wider period of time. In other words, the more layers the model has, the further into the past it can "see". That receptive field $t_{rf}$ can be described as $t_{rf}=4n_b(2^n_c-1) + 1$ with $n_b$ as the number of dilation blocks, and $n_c$ as the number of cells in each block.

\subsection{The Pipeline}
The pipeline has two chained modules. Module A predicts future instances' function names. Module B predicts the arrival time of those instances. This design allows more flexibility in policy integration and continuous operation/backup planning. In particular, when the main model for module A has problems, the backup model for module A may seamlessly take over and produce the instances' function names needed by module B. Module A and B forecasts can power different yet complimentary policy sub-sets to be discussed in later sections.

\begin{figure*}[htbp]
\centerline{\includegraphics[width=5.6in]{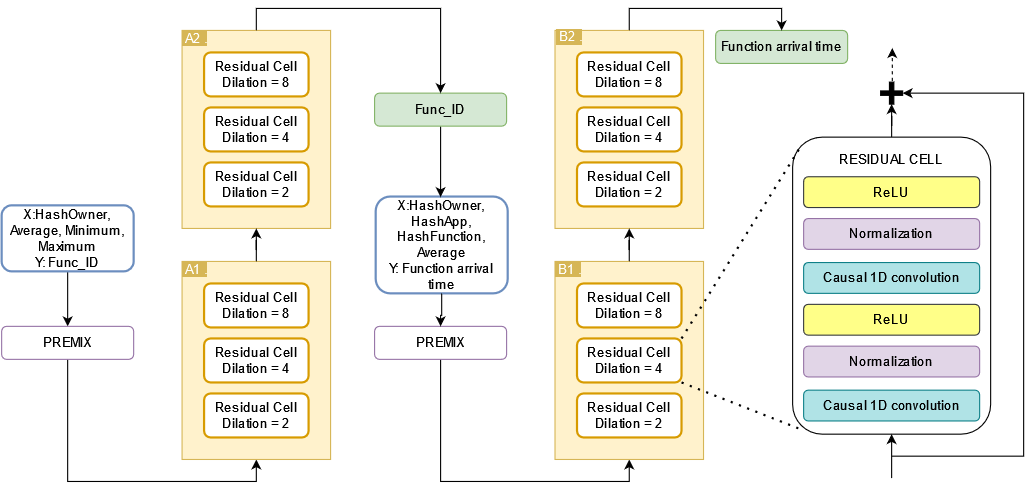}} 
\caption{The Pipeline Design}
\label{fig_modelDesign}
\end{figure*}

Each module has one or two blocks. This number of blocks represents the depth of the TCN pipeline. Each block contains three types of residual cells with exponentially increased dilation from 2. In each residual cell, we have two stacked sets of the causal convolution layer similar to figure \ref{fig_dilation}, followed by a normalization, and a nonlinear activation (ReLU). The number of cells in each block ranges from 3 to 6 cells. The number of hidden channels ranges from 64 to 256. The final output of a residual cell is a combination of its stacks' output and its initial input. Figure \ref{fig_modelDesign} visually summarizes the pipeline.

\begin{figure}[htbp]
\centerline{\includegraphics[width=3.3in]{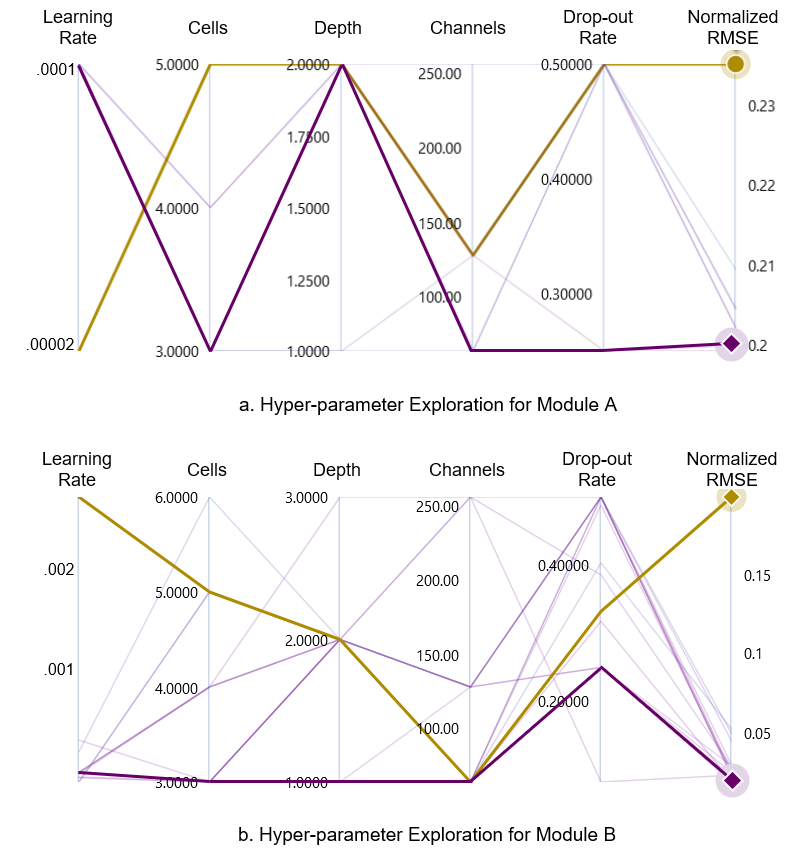}} 
\caption{Automatic Hyper-parameter Exploration Results}
\label{fig_paramExplore}
\end{figure}

Auto ML strategy was employed to probe for the best hyper-parameter set. For each module, 12 landmark TCN models were built with hyper-parameter sets that reasonably span the hyper-parameter space. Each model's performance was measured. The best performing landmark models' hyper-parameter sets were then used to calculate fine-tuned hyper-parameter sets for locating the best performing model. The model exploration time budget was set to a max of 80 hours. This is due to Azure cloud personal use's resource constraints. Finally, the best performing model was benchmarked against other ML-based forecasting models on the same dataset. Figure \ref{fig_paramExplore} shows the results of automatic hyper-parameter explorations for both module A (predicting HTTP-triggered instances' function names) and module B (predicting HTTP-triggered instances' arrival time).

Relying on these results, future developers and researchers who want to adopt this work only need to explore around the provided best performing TCN hyper-parameter sets. The specific performance results of the best performing TCN models are further discussed in the next section.

\section{The Results}\label{sec_results}
Models for both modules were evaluated on a full range of metrics suitable for time series forecasting task. The models were evaluated at each cross validation step which equals the forecast horizon. For brevity, the paper only lists the evaluation results at the last cross validation step. The full evaluation results of all models for both modules can be found at \textit{https://github.com/Cybonto/TCN4CLOUD}. The key metrics are as follows.

Explained variance represents the percent decrease in variance of the original data to the variance of the errors. The closer the explained variance is to 1 the better. Mean absolute percentage error (MAPE) measures the average difference between a predicted value and the actual value. The closer MAPE is to 0 the better. Root mean squared error (RMSE) equals to the square root of the squared difference between a predicted value and the actual value. A normalized RMSE is the RMSE divided by the range of the data. The closer a normalized RMSE is to 0 the better. R2 score describes the decreased mean square error with respect to the total data variance. The closer R2 is to 1 the better. Spearman correlation measures the monotonicity of relationship between two datasets. The closer a Spearman score is to 1 the better. Table \ref{tab:task1_perf} provides the benchmark results for predicting instances' function names (module A), and table \ref{tab:task2_perf} provides the benchmark results for predicting instances' arrival time (module B). 

\begin{table}[]
\caption{Predicting Instances' Function Names Performance}
\label{tab:task1_perf}
\begin{tabular}{|l|r|r|r|r|}
\hline
Metric & TCN & Arima & ES$^{\mathrm{b}}$ & Prophet \\ \hline
Explained variance & -.002 & -.002 & $\sim$0 & .012 \\ \hline
Mean absolute percentage error & 1968 & 136 & 126 & 142 \\ \hline
Normalized root mean square error & .200 & .238 & .286 & .220 \\ \hline
R2 score & -0.05 & -0.29 & -0.62 & -0.12 \\ \hline
Spearman Correlation & -.042 & .078 & -1 & 0.15 \\
\hline
\multicolumn{5}{l}{$^{\mathrm{a}}$ RMSE: Root Mean Square Error.}\\
\multicolumn{5}{l}{$^{\mathrm{b}}$ ES: Exponential Smoothing.}
\end{tabular}
\end{table}

\begin{table}[]
\caption{Predicting Instance Arrival Time Performance}
\label{tab:task2_perf}
\resizebox{\columnwidth}{!}{%
\begin{tabular}{|l|r|r|r|r|r|}
\hline
Metric & TCN & Arimax & Arima & ES$^{\mathrm{b}}$ & Prophet \\ \hline
Explained variance & .50 & .32 & .20 & $\sim$0 & -.002 \\ \hline
Mean absolute percentage error & 15.79 & 230 & 210 & 291 & 252 \\ \hline
Normalized RMSE$^{\mathrm{a}}$ & .16 & .18 & .20 & .20 & .27 \\ \hline
R2 score & -.87 & -.39 & -.59 & -.80 & -.80 \\ \hline
Spearman Correlation & .92 & .32 & .32 & -1 & -.05 \\
\hline
\multicolumn{6}{l}{$^{\mathrm{a}}$ RMSE: Root Mean Square Error}\\
\multicolumn{6}{l}{$^{\mathrm{b}}$ ES: Exponential Smoothing}
\end{tabular}%
}
\end{table}

\subsection{The Proposed Ensemble Policy}
\begin{figure*}[t!]
\centerline{\includegraphics[width=5.7in]{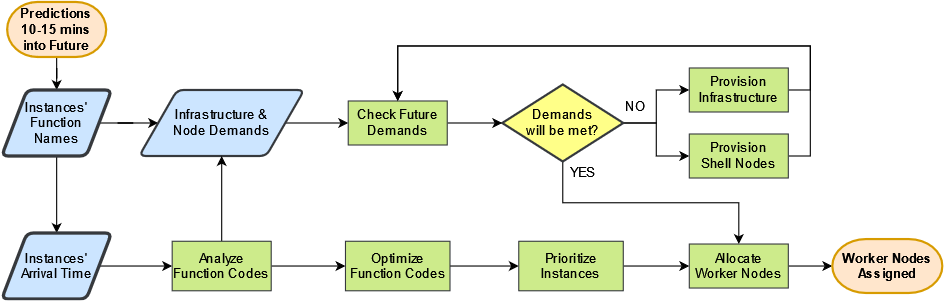}} 
\caption{The Ensemble Cold-start Management Policy}
\label{fig_policy}
\end{figure*}

Predicting what new function instance will be invoked (module A) is much harder than predicting when those functions will be invoked (module B). For the main reason, the need for creating a new function instance depends on both the FaaS environment states and the FaaS users' business demands. The later's information may not be embedded in past instance provisions. Therefore, the cost of building a TCN model for module A can be significant if the number of registered functions is high. In this study, the MS Azure ML default compute quota was not not enough to prevent Module A's TCN model training crash when the models have a depth of 2 and more than 150 hidden channels. Therefore, the results shown in table \ref{tab:task1_perf} are those of the best landmark model instead of a fine-tuned model for module A. Even so, the landmark model's performance is competitive with the best non-deep-learning model (Prophet). Companies with more resource might be able to build a much better performing TCN model using the same codes but with different hyper-parameters. 

For predicting the instances' arrival time when future instances' function names are known, TCN outperform non-deep-learning models as shown in table \ref{tab:task2_perf}. TCN training for this Module B task on MS Azure ML was possible because the hyper-parameters for the best performing model only require the depth of 1 and less than 80 hidden channels.

\section{Proposed Policy}\label{sec_policy}

Solving the cloud serverless cold-start problem is not trivial as the root causes cover different levels of the serverless stack. Therefore, this paper proposes an ensemble policy that relies on the strategies of divide-and-conquer and parallel processing. Based on the previous works, the paper categorizes cold-start sub-problems into two sets of infrastructure-level problems and function-level problems. Assuming the TCN models reliably give predictions of 10 to 15 minutes into the future as previously discussed, we should have enough time to ensemble various cold-start management strategies, orchestrate, and automate them. Predicted function names can be used to solve infrastructure-level problems while predicted function arrival times can be used to solve function-level problems. Because of the TCN pipeline's loose coupling and tight cohesion design, the papers' proposed policy can address some of the problems at both levels at the same time. The specific details are as follows. 

\subsection{Previous Works}

Previous works showed that the setup time of worker nodes and related supporting infrastructure is expensive \cite{Singhvi2021Atoll:Platform, Mohan2019AgileServerless}. The solutions include preemptively provision network cache and shell-nodes \cite{Mohan2019AgileServerless}, preemptively provision new shell nodes and load-balance existing worker nodes across physical clusters \cite{Singhvi2021Atoll:Platform}, rank and select servers for function scheduling \cite{Yu2021FaaSRank:Platforms}, and preemptively provision the worker nodes based on node cache miss rate \cite{Fuerst2021FaasCache:Caching}. Keep-alive policies can be based on operational priorities such as cost, frequencies, time, size \cite{Fuerst2021FaasCache:Caching}.

At the function-level, function execution priorities are unclear from both customer perspective (i.e. customers do not specify function priorities while registering their functions) and CSP perspective (i.e. functions were not optimally scheduled to minimize costs while satisfying execution bounds) \cite{Singhvi2021Atoll:Platform}. The solutions involve analyzing dependencies, dead-lines \cite{Singhvi2021Atoll:Platform}, environment states, actions, rewards, and other metrics to preemptively provision and terminate function-supporting resources in the right order \cite{Agarwal2021AFrequency, Jegannathan2022APlatform, Bermbach2020UsingServices}. Specifically, a model can set functions' pre-warm time and keep-alive time based on dependency-predictable functions' invocation histograms, while using fixed-timeout policy for dependency-unpredictable functions \cite{Shen2021Defuse:Platforms}. Function-level cold-start problems can also be manged by reducing application code loading latency by loading critical functions first and optional functions later \cite{Wen2022LambdaLite:Computing}, or by fusing and optimizing function codes for more efficient run-time execution \cite{Schirmer2022Fusionize:Fusion, Lee2021MitigatingFusion}.

\subsection{The Ensemble Policy}

Because the cold-start problem has roots in several serverless cloud layers, it is best to attack the problem simultaneously from multiple angles. Anchoring on TCN's capability to predict function instance arrivals 10 to 15 minutes into the future, this paper proposes an ensemble policy that utilizes relevant previously mentioned works. In particular, the policy has two parallel paths (A and B) that can be described as follows.

In path A, after receiving the predicted instances' function names from module A model, future (next 10 to 15 minutes) infrastructure and worker node demands can be inferred. This is possible because all functions should be previously registered, allowing coarse-grain mappings of function names with system requirements. Finer-grain details can be provided by path B's in-depth analysis of function codes. If future demands will not be met, more infrastructure supports and/or shell nodes will be preemptively provisioned per \cite{Mohan2019AgileServerless, Singhvi2021Atoll:Platform, Agarwal2021AFrequency, Jegannathan2022APlatform, Bermbach2020UsingServices}. If future demands can be met, the system will rank existing available resources and orchestrate them per \cite{Yu2021FaaSRank:Platforms}. Resource ranking and infra-structure orchestration schedules will be shared with path B.

In path B, codes of predicted function instances will be analyzed based on the predicted arrival time. This can be at disk i/o read speed assuming code analysis was performed when customers registered the functions with the CSP. Finer-grain code dependencies can be inferred and pass to path A to assist with future demand inferences. Function codes can then be optimized per \cite{Schirmer2022Fusionize:Fusion, Lee2021MitigatingFusion}. Based on function instances' arrival time, fine-grain requirements, and optimized function code graphs, instances can be prioritized. For example, the instance that was predicted to arrive first may not have the required resources provisioned first. In the last step, path B combines all the calculated information it has together with path A's resource ranking and infra-structure orchestration schedules to decide which shell nodes will be turned into which worker nodes. Figure \ref{fig_policy} summarized the policy.

\section{Conclusion}
Serverless cloud computing is an attractive cloud service model as it offers businesses the same cloud advantages at a fraction of other cloud service models' price. One of the biggest problem with the serverless cloud is the paradox of shutting down idling cloud resources to keep the costs low versus keeping the idling resources available for prompt executions of customers' requests. If there is no readily available cloud resources to handle customers' requests, new resources must be provisioned. The cold-start problem happens when provisioning process is unacceptably long. State of the art solutions address the cold-start problem from different angles at different serverless cloud layers. The gaps mostly involve unrealistic assumptions/simulations about the serverless cloud workloads, and the narrow focus.

This paper has a more holistic approach to the cold-start problem. First, a set of real-life challenges within the serverless cloud environment was identified. This set is then operationalized into data set requirements and predictive model requirements. Based on these requirements, a Temporal Convolutional Network model was implemented, allowing predictions of function instances and their arrivals 10 to 15 minutes into the future. On top of this capability, an ensemble policy was constructed. This policy embraces parallelism, low-coupling high-cohesion principles, and utilizes relevant previous works to address the cold-start problem at both infrastructure and function levels. Going beyond cold-start management, the proposed policy and publicly available codes can be adopted in solving other cloud problems such as optimizing the provisioning of virtual software-defined network assets.

\bibliographystyle{IEEEtran}
\bibliography{IEEEabrv,references.bib}

\begin{thebibliography}{10}
\providecommand{\url}[1]{#1}
\csname url@samestyle\endcsname
\providecommand{\newblock}{\relax}
\providecommand{\bibinfo}[2]{#2}
\providecommand{\BIBentrySTDinterwordspacing}{\spaceskip=0pt\relax}
\providecommand{\BIBentryALTinterwordstretchfactor}{4}
\providecommand{\BIBentryALTinterwordspacing}{\spaceskip=\fontdimen2\font plus
\BIBentryALTinterwordstretchfactor\fontdimen3\font minus
  \fontdimen4\font\relax}
\providecommand{\BIBforeignlanguage}[2]{{%
\expandafter\ifx\csname l@#1\endcsname\relax
\typeout{** WARNING: IEEEtran.bst: No hyphenation pattern has been}%
\typeout{** loaded for the language `#1'. Using the pattern for}%
\typeout{** the default language instead.}%
\else
\language=\csname l@#1\endcsname
\fi
#2}}
\providecommand{\BIBdecl}{\relax}
\BIBdecl

\bibitem{Shahrad2020ServerlessProvider}
M.~Shahrad, R.~Fonseca, I.~Goiri, G.~Chaudhry, P.~Batum, J.~Cooke, E.~Laureano,
  C.~Tresness, M.~Russinovich, and R.~Bianchini, ``{Serverless in the wild:
  Characterizing and optimizing the serverless workload at a large cloud
  provider},'' \emph{Proceedings of the 2020 USENIX Annual Technical
  Conference, ATC 2020}, pp. 205--218, 2020.

\bibitem{Shafiei2022ServerlessApplications}
H.~Shafiei, A.~Khonsari, and P.~Mousavi, ``{Serverless Computing: A Survey of
  Opportunities, Challenges, and Applications},'' \emph{ACM Computing Surveys},
  vol.~54, no. 11s, pp. 1--32, 2022.

\bibitem{Tayal2019DeterminingConsulting}
A.~Tayal, E.~Lam, D.~Choudhury, M.~Dickerson, G.~Moovera, and G.~Arora,
  ``{Determining the Total Cost of Ownership of Serverless Technologies when
  compared to Traditional Cloud v2 Deloitte Consulting},'' Tech. Rep.
  September, 2019.

\bibitem{Mahmoudi2020PerformancePlatforms}
N.~Mahmoudi and H.~Khazaei, ``{Performance Modeling of Serverless Computing
  Platforms},'' \emph{IEEE Transactions on Cloud Computing}, vol.~10, no.~4,
  pp. 2834--2847, 2020.

\bibitem{Ebrahimpour2022AEnvironments}
H.~Ebrahimpour, ``{A heuristic-based package-aware function scheduling approach
  for creating a trade-off between cold-start time and cost in FaaS computing
  environments},'' 2022.

\bibitem{Jegannathan2022APlatform}
A.~P. Jegannathan, R.~Saha, and S.~K. Addya, ``{A Time Series Forecasting
  Approach to Minimize Cold Start Time in Cloud-Serverless Platform},''
  \emph{2022 IEEE International Black Sea Conference on Communications and
  Networking, BlackSeaCom 2022}, pp. 325--330, 2022.

\bibitem{Steinbach2022TppFaaS:Processes}
M.~Steinbach, A.~Jindal, M.~Chadha, M.~Gerndt, and S.~Benedict, ``{TppFaaS:
  Modeling Serverless Functions Invocations via Temporal Point Processes},''
  \emph{IEEE Access}, vol.~10, pp. 9059--9084, 2022.

\bibitem{Singhvi2021Atoll:Platform}
A.~Singhvi, A.~Balasubramanian, K.~Houck, M.~D. Shaikh, S.~Venkataraman, and
  A.~Akella, ``{Atoll: A scalable low-latency serverless platform},''
  \emph{SoCC 2021 - Proceedings of the 2021 ACM Symposium on Cloud Computing},
  pp. 138--152, 2021.

\bibitem{Lin2021ModelingApplications}
C.~Lin and H.~Khazaei, ``{Modeling and Optimization of Performance and Cost of
  Serverless Applications},'' \emph{IEEE Transactions on Parallel and
  Distributed Systems}, vol.~32, no.~3, pp. 615--632, 2021.

\bibitem{Das2020PerformancePlacement}
A.~Das, S.~Imai, S.~Patterson, and M.~P. Wittie, ``{Performance Optimization
  for Edge-Cloud Serverless Platforms via Dynamic Task Placement},''
  \emph{Proceedings - 20th IEEE/ACM International Symposium on Cluster, Cloud
  and Internet Computing, CCGRID 2020}, no.~1, pp. 41--50, 2020.

\bibitem{Mampage2022ADirections}
A.~Mampage, S.~Karunasekera, and R.~Buyya, ``{A Holistic View on Resource
  Management in Serverless Computing Environments: Taxonomy and Future
  Directions},'' \emph{ACM Computing Surveys}, vol.~54, no. 11s, pp. 1--36,
  2022.

\bibitem{Agarwal2021AFrequency}
S.~Agarwal, M.~A. Rodriguez, and R.~Buyya, ``{A reinforcement learning approach
  to reduce serverless function cold start frequency},'' \emph{Proceedings -
  21st IEEE/ACM International Symposium on Cluster, Cloud and Internet
  Computing, CCGrid 2021}, pp. 797--803, 2021.

\bibitem{Kaffes2022Hermod:Functions}
K.~Kaffes, N.~J. Yadwadkar, and C.~Kozyrakis, ``{Hermod: Principled and
  Practical Scheduling for Serverless Functions},'' \emph{SoCC 2022 -
  Proceedings of the 13th Symposium on Cloud Computing}, pp. 289--305, 2022.

\bibitem{Benidis2022DeepSurvey}
K.~Benidis, S.~S. Rangapuram, V.~Flunkert, Y.~Wang, D.~Maddix, C.~Turkmen,
  J.~Gasthaus, M.~Bohlke-Schneider, D.~Salinas, L.~Stella, F.~X. Aubet,
  L.~Callot, and T.~Januschowski, ``{Deep Learning for Time Series Forecasting:
  Tutorial and Literature Survey},'' \emph{ACM Computing Surveys}, vol.~55,
  no.~6, 2022.

\bibitem{Mohan2019AgileServerless}
A.~Mohan, H.~Sane, K.~Doshi, S.~Edupuganti, N.~Nayak, and V.~Sukhomlinov,
  ``{Agile cold starts for scalable serverless},'' \emph{11th USENIX Workshop
  on Hot Topics in Cloud Computing, HotCloud 2019, co-located with USENIX ATC
  2019}, 2019.

\bibitem{Yu2021FaaSRank:Platforms}
H.~Yu, A.~A. Irissappane, H.~Wang, and W.~J. Lloyd, ``{FaaSRank: Learning to
  Schedule Functions in Serverless Platforms},'' \emph{Proceedings - 2021 IEEE
  International Conference on Autonomic Computing and Self-Organizing Systems,
  ACSOS 2021}, pp. 31--40, 2021.

\bibitem{Fuerst2021FaasCache:Caching}
A.~Fuerst and P.~Sharma, ``{FaasCache: Keeping serverless computing alive with
  greedy-dual caching},'' \emph{International Conference on Architectural
  Support for Programming Languages and Operating Systems - ASPLOS}, pp.
  386--400, 2021.

\bibitem{Bermbach2020UsingServices}
D.~Bermbach, A.~S. Karakaya, and S.~Buchholz, ``{Using application knowledge to
  reduce cold starts in FaaS services},'' \emph{Proceedings of the ACM
  Symposium on Applied Computing}, pp. 134--143, 2020.

\bibitem{Shen2021Defuse:Platforms}
J.~Shen, T.~Yang, Y.~Su, Y.~Zhou, and M.~R. Lyu, ``{Defuse: A dependency-guided
  function scheduler to mitigate cold starts on faas platforms},''
  \emph{Proceedings - International Conference on Distributed Computing
  Systems}, vol. 2021-July, pp. 194--204, 2021.

\bibitem{Wen2022LambdaLite:Computing}
\BIBentryALTinterwordspacing
J.~Wen, Z.~Chen, D.~Li, J.~Chen, Y.~Liu, H.~Wang, X.~Jin, and X.~Liu,
  ``{LambdaLite: Application-Level Optimization for Cold Start Latency in
  Serverless Computing},'' 2022. [Online]. Available:
  \url{http://arxiv.org/abs/2207.08175}
\BIBentrySTDinterwordspacing

\bibitem{Schirmer2022Fusionize:Fusion}
T.~Schirmer, J.~Scheuner, T.~Pfandzelter, and D.~Bermbach, ``{Fusionize:
  Improving Serverless Application Performance through Feedback-Driven Function
  Fusion},'' \emph{Proceedings - 2022 IEEE International Conference on Cloud
  Engineering, IC2E 2022}, pp. 85--95, 2022.

\bibitem{Lee2021MitigatingFusion}
S.~Lee, D.~Yoon, S.~Yeo, and S.~Oh, ``{Mitigating cold start problem in
  serverless computing with function fusion},'' \emph{Sensors}, vol.~21,
  no.~24, 2021.

\end{thebibliography}

\end{document}